# 3D Spectroscopic tracking of individual Brownian nanoparticles during galvanic exchange

Minh-Chau Nguyen,[1,2] Pascal Berto,[2,3] Fabrice Valentino,[3] Jean-François Lemineur,[1] Jean-Marc Noel,[1] Frédéric Kanoufi,[1,*] Gilles Tessier[2,3,*]

[1] Université de Paris Cité, ITODYS, CNRS, 75006 Paris, France

[2] Sorbonne Université, INSERM, CNRS, Institut de la Vision, 17 rue Moreau, F-75012, Paris, France

[3] Université Paris Descartes, CNRS, UMR 8250, 45 rue des Saints-Pères, F-75006 Paris, France

**\* corresponding author :** **frederic.kanoufi@u-paris.fr** and **gilles.tessier@sorbonne-universite.fr**

ABSTRACT. Monitoring chemical reactions in solutions at the scale of individual entities is challenging: single particle detection requires small confocal volumes which are hardly compatible with Brownian motion, particularly when long integration times are necessary. Here, we propose a real-time (10 Hz) holography-based nm-precision 3D tracking of single moving nanoparticles. Using this localization, the confocal collection volume is dynamically adjusted to follow the moving nanoparticle and allow continuous spectroscopic monitoring. This concept is applied to the study galvanic exchange in freely-moving colloïdal silver nanoparticles with gold ions generated in-situ. While the Brownian trajectory reveals particle size, spectral shifts dynamically reveal composition changes and transformation kinetics at the single object level, pointing at different transformation kinetics for free and tethered particles.

KEYWORDS. Spectroscopy – Electrochemistry – Brownian nanoparticle – Galvanic exchange - Holography.

**INTRODUCTION**

Unlike molecules, most nanostructured materials, e.g. synthetic batches of NanoParticles (NPs), typically include a broad diversity of individual objects in terms of size, morphology, surface chemistry, atomic arrangement, etc. Revealing intrinsic structure-activity relationships within such diverse assemblies is essentially impossible using conventional ensemble-averaged measurements. A detailed and accurate understanding of the chemical performances of nanostructured materials thus calls for individual, nanoscale probing of chemical reactions. Various strategies have been proposed to visualize and quantify *in situ* chemical reactions from the single-NP- to the single-active-site-level.

Among these strategies, several optical microscopy configurations able to probe *in situ* and in real time electrochemical or chemical processes with single nanoobject sensitivity have been recently reviewed.[1–15] However, the optically inspected NPs are mostly static objects immobilized on a surface or in a gel-like matrix, or fabricated using top-down approaches. Immobility is indeed essential to provide sufficient instrument integration time when imaging faint objects.[16–21] A wide range of chemical reactions at the single NPs level have been reported, ranging from conformational change of the NP capping agent, which can be used as a sensing platform,[8,10,22] to the electrochemical or chemical conversion of the NP composition.[7–9,23,24] However, such measurements are often performed at surface-tethered NPs, removing the complexity of bulk vs interfacial chemical processes. While particularly useful to study interfacial chemistry, or electrochemistry, these techniques cannot be applied to the most important type of solution-phase nanochemistries, which involve Brownian NPs, freely diffusing in reactive solutions.

Tracking chemically-induced transformations of the shape and composition of individual NPs in solutions is essential to various analytical purposes. Among them, electroanalytical strategies, such as the electrochemical nanoimpact strategy, have been developed to study the intrinsic activity of NPs. It relies on the transport of NPs from the solution to an electrode where their electrochemical activity can be probed individually. In these methods, a simultaneous NP visualization or localization by optical microscopy is essential in order to provide mechanistic insights which can be correlated to the electrochemistry.[25–32] From transport observation, single object hydrodynamic size is often measured, via Nanoparticle Tracking Analysis (NTA) strategies, enabling to monitor NP size increase (during e.g. polymer, metal or oxide NP synthesis) or dissolution. However, whenever a chemical composition change is involved, a complementary spectroscopic identification is essential. UV-vis spectroscopy is commonly used to probe the dynamics of chemical transformations in NPs ensembles, particularly in the case of plasmonic NPs.[9,23,33–36]

Galvanic exchange[37–39] is one such reaction which can be efficiently probed at the single NP level using dark-field spectroscopic microscopy. It is a redox reaction between a metallic NP (e.g. Ag) and a solution of a metal ion (more noble, e.g. $Au^{3+}$) leading, as in eq (1),

$$3Ag_{(s)} + Au^{3+}_{(aq)} \rightarrow 3Ag^{+}_{(aq)} + Au_{(s)} \qquad (1)$$

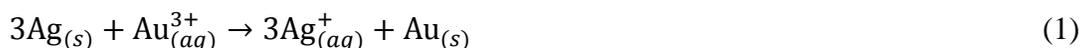

to the replacement of the metal atoms of the NP, during its oxidation, by atoms of the more noble metal then reduced (more positive standard redox potential). Such reactions provide valuable pathways to construct NPs with a rich variety of chemical compositions or architectures, e.g. hollow[40–42] or core/shell nanostructures. In the model situation of an Ag NP, the reaction (1) yields the replacement of the outer Ag atoms of the NPs by Au atoms, resulting in an Ag@Au core-shell NP.[43–45] Beyond model systems, Cu, Co or Ni NPs have also been exchanged with more noble metals (Au, Pt, Pd and Ag).[46–49] Such synthesis routes have also been more recently extended to metal oxides and other ionic nanocrystals.[50] Ion exchange reactions draw attention as a simple, versatile solution-phase chemistry method to transform NPs into nanostructures with diverse complex chemical compositions and architectures. Even if the galvanic replacement reaction is ruled by the difference in standard redox potentials, $E^0$, of the different metals, e.g. for the model deposition of Au ($E^0$ = 1.50V vs Standard Hydrogen Electrode, SHE),[37,42] reactions in Ag NPs ($E^0$ = 0.80V vs SHE) proceed through intricate solid phase transformation mechanisms. In this respect, different groups have demonstrated the importance of exploring such mechanisms at the single NP level rather than from ensemble-averaged approaches. High-resolution electron microscopy imaging allows to identify the various intermediate nanostructures produced, starting from the formation of voids.[51–53] Since the size, composition, or shape modifications occurring during metal exchange strongly affect the NPs optical properties, particularly through the surface plasmon resonance of plasmonic NPs, optical techniques allow the sensitive and dynamic monitoring of such transformations. At the single NP level, Jain's group proposed an in situ plasmonic scattering spectroscopic microscopy monitoring, showing, even in the simplest model system, that the ion exchange could proceed from a fast NP transformation with cooperative NP-to-NP propagation[1,16,54] to a slower and non-cooperative transformation.[54]

Like most single entity spectroscopic studies, these characterizations were restricted to individual NPs immobilized on surfaces in order to reach spectroscopic integration times allowing single-particle scale sensitivity.[16–18] The most relevant metal exchange reactions, however, are often realized in colloidal solutions, and static studies may partially hinder or modify the transformation symmetries available in a solution phase, notably due to slower renewal of reagents near the NP-electrode interface.[37,42,55,56] Yet, single-NP spectroscopy in solutions is quite challenging, since acquiring spectra on individual NPs with weak scattering cross-sections requires long integration times, and the few available photons are dispersed over several pixels of the spectrograph. In addition, single-particle spectroscopy demands spatial selectivity, typically confocal, to exclude spurious signals from other objects in the solution. When using confocal detection, however, a particle in Brownian motion will move away from the position where the light it scatters can be collected by the spectrometer. Most NPs leave the confocal collection region within a much shorter time than what is necessary to a spectral acquisition. So far, the main response to these strong instrumental constraints has been either to study ensembles (to increase the available optical signal) or to immobilize the NPs (to increase integration times). Here, we propose an approach enabling long confocal acquisitions on moving single objects, based on the use of a real-time adaptive optical system to compensate the random 3D motion of the NP: the confocal collection volume of the spectrometer is moved in real time to follow the particle to any x,y,z position in the sample.

In this work, we propose i) a method to monitor individual freely moving NPs both spectrally and spatially, ii) a technique for the controlled release of gold ions at high dilution, providing well-controlled galvanic exchange conditions in order to allow iii) measurements able to elucidate the transformation scenario, which point at distinct transformation kinetics for freely moving and tethered NPs.

To determine the particle position and the necessary correction, we propose here to use digital holographic microscopy[57–60] to obtain a sensitive (down to 50 nm diameter on noble metal NPs) and accurate (nm-range precision) 3D localization of the NP over a broad axial range (typ. 100 µm, mostly limited by the axial aberrations of the microscope objective). The NP position is computed in real-time using a graphics processing unit (GPU), to drive a 3-coordinates adaptive optics system (2 transversal axes plus longitudinal refocusing) which moves the confocal collection volume of a spectrometer to follow the NP. This real-time correction allows spectroscopic acquisitions of any duration, even on particles moving in a microfluidic chamber, as depicted in Figure 1, thus allowing the spectral monitoring of single NP in Brownian motion.

This system thus provides access to two simultaneous and orthogonal NP characterizations: the 3D trajectory of the NP gives access to its hydrodynamic size via its mean square displacement, while spectroscopy gives an insight into its composition and geometry. Here, an electrode is used to deliver $Au^{3+}$ ions in a microfluidic chamber, providing temporal and spatial control of the concentration in the solution, as well as a fine control of the metal exchange reaction kinetics. The coupled use of real-time tracking and spectroscopy on chosen NPs provides insight into the dynamics of size and composition changes during the galvanic modification of individual NPs.

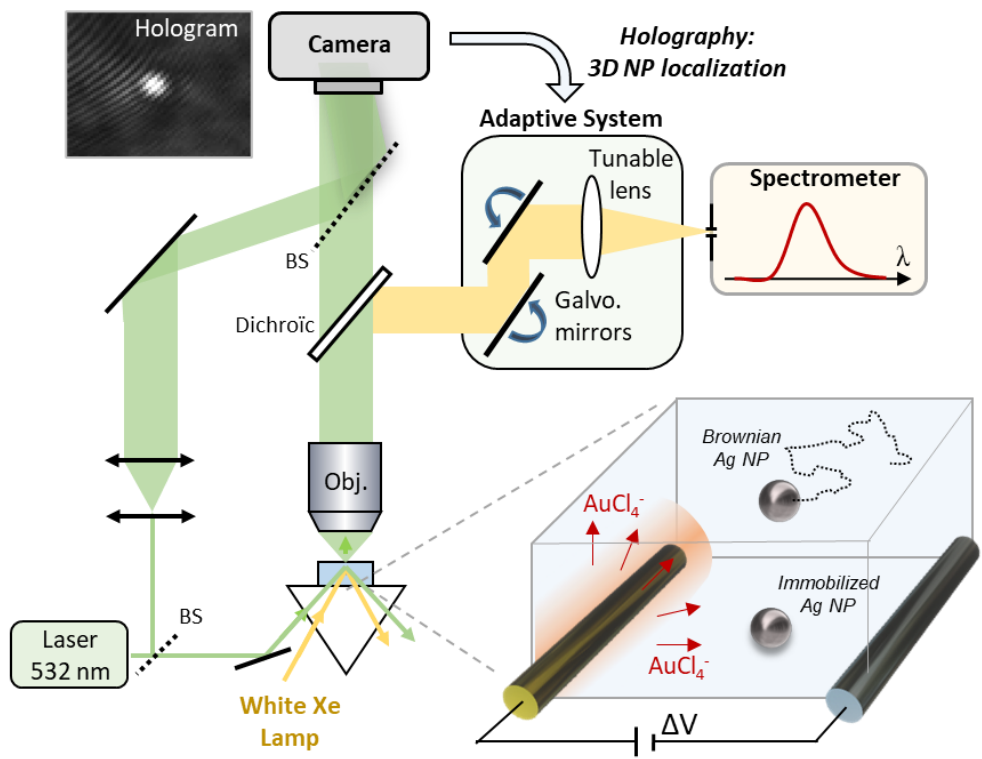

**Figure 1**: Experimental setup combining holographic microscopy and adaptive spectroscopy to monitor the electrochemically-triggered galvanic exchange of single Ag nanoparticle, NP, with electrogenerated $AuCl_4^-$ ions. The NP is illuminated by a white Xe arc lamp, and by a $\lambda=532$ nm laser. This laser light is used to determine the 3D coordinates of the NP in real-time by holography. This 3D position drives an adaptive optical system (galvanometric mirrors and tunable lens) which moves the confocal collection volume of a spectrometer to allow the collection and spectral analysis of the light scattered by fixed or freely moving Brownian NPs. In the microfluidic chamber, a sacrificial Au electrode releases $AuCl_4^-$ at electrochemically controlled flux which diffuse towards Ag NPs (60 or 100 nm).

## RESULTS / DISCUSSION

**Electrogeneration as a means to control the galvanic exchange dynamics**

To allow 3D optical visualization, the experiments were carried out in a parafilm-sealed ca. 1cmx1cmx500µm microfluidic electrochemical chamber filled with a 25mM HCl solution (see

Methods/Experimental section). The chamber contained a 2-electrode electrochemical cell made of a Au microwire (ca. 5mm long, 250μm diameter) and a Ag/AgCl wire used respectively as the working electrode (WE) and the counter electrode (CE).

The kinetics of galvanic exchange reactions operating in NPs depends on the solid phase transformation and the solution chemistry surrounding the NP. It is a rather fast process, typically completed within milliseconds, as observed in ensemble-averaged experiments.[1,16] In order to observe and sample properly the evolution of this reaction using an optical microscope read-out at the single NP level, the reaction kinetics must be slowed down to second-range timescales. This can be achieved by using μM dilution of the replacing ions solution, which can then be delivered over NPs immobilized on an inert surface using microfluidics. However, when using freely moving particles, the injection of liquid to trigger the galvanic exchange reaction generates relatively strong fluid flows. This displaces NPs out of the limited Field of View (FoV) of the microscope, typ. 50μmx50μm, and forbids a complete investigation of NPs transformation. A possible workaround is to generate the gold ions *in situ* by anodic dissolution of a gold wire.[61] This dissolution is facilitated in HCl solution, according to the overall electrode reaction:

$$\text{Au} + 4\text{Cl}^- \rightarrow \text{AuCl}_4^- + 3\text{e}^- \qquad (2)$$

and with a standard reduction potential of 1.002 V vs Standard Hydrogen Electrode (SHE), it occurs before the solvent oxidation discharge (oxygen or chlorine evolution reactions).

When a potential difference of 1.1 V is applied between the Au WE and the Ag/AgCl CE, an electrochemical current flows in the circuit, resulting in the electrodissolution of the Au electrode and electrogeneration of $AuCl_4^-$ ions. The concentration of $AuCl_4^-$ ions at the electrode surface is driven by the current flowing at the electrode. As schematized in Figure 1, the ions then radially diffuse into the solution towards Ag NPs, and their conversion is initiated. This conversion rate then depends on the $AuCl_4^-$ concentration profile in the electrochemical chamber, which varies both with time and distance to the Au wire (here, $d \sim$ 1 to 2 mm). A more detailed description and a Comsol® modeling of the spatio-temporal expansion of the $AuCl_4^-$ ions concentration in this experimental geometry are given in Section SI 1.

Typically, for the $d_{NP}$=60nm or 100nm NPs studied here, µM-range $[AuCl_4^-]_{loc}$ concentrations correspond to fluxes $f$ of the order of $3 \cdot 10^5$ and $6 \cdot 10^5$ $AuCl_4^-$ per second respectively. This is the order of magnitude of the number of Ag atoms in the first outer atomic layer of these Ag NPs, suggesting that through this electrochemical triggering, the monitoring of submonolayer Ag NP transformation dynamics is within reach.

**Monitoring galvanic exchange at surface-immobilized single Ag NPs**

The dilute electrogenerated $AuCl_4^-$ ions diffusing away from the Au electrode will react with the Ag NPs. In the presence of chloride-based electrolyte, the oxidation products of the Ag NP can be $Ag^+$ and AgCl, and the reaction scheme is expected to proceed according to:

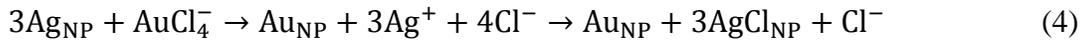

$$3Ag_{NP} + AuCl_4^- \rightarrow Au_{NP} + 3Ag^+ + 4Cl^- \rightarrow Au_{NP} + 3AgCl_{NP} + Cl^- \qquad (4)$$

The formation of either $Ag^+$ or AgCl depends on the rate of the galvanic exchange reaction and therefore on the local $Ag^+$ concentration (saturation) surrounding each individual Ag NP. Typically, from the 10mM HCl electrolyte and the AgCl solubility product ($1.8 \cdot 10^{-10}$ $M^{-2}$), the saturation of $Ag^+$ is expected to be 18nM, suggesting that the neighborhood of the NP is oversaturated in $Ag^+$ ions, and that AgCl is precipitating on the NP. The nanoparticle is then expected to contain three different inclusions: Ag, Au, and AgCl,[16,54] as previously observed by Jain et al using dark-field spectroscopic microscopy on single NP immobilized on an inert surface.

In order to validate the 3D optical spectroscopy probing of galvanic exchange at Brownian Ag NPs, experiments were first carried out on surface-immobilized NPs. They were deposited on a glass coverslip by dropcasting a droplet of an aqueous dispersion of either 60 nm or 100 nm diameter Ag NPs. After evaporation of the solvent, the coverslip was used as the bottom part of the microfluidic cell. The FoV was selected in a region displaying individual Ag NPs separated by distances well above the optical resolution limit, > 10μm from any nearest NP.

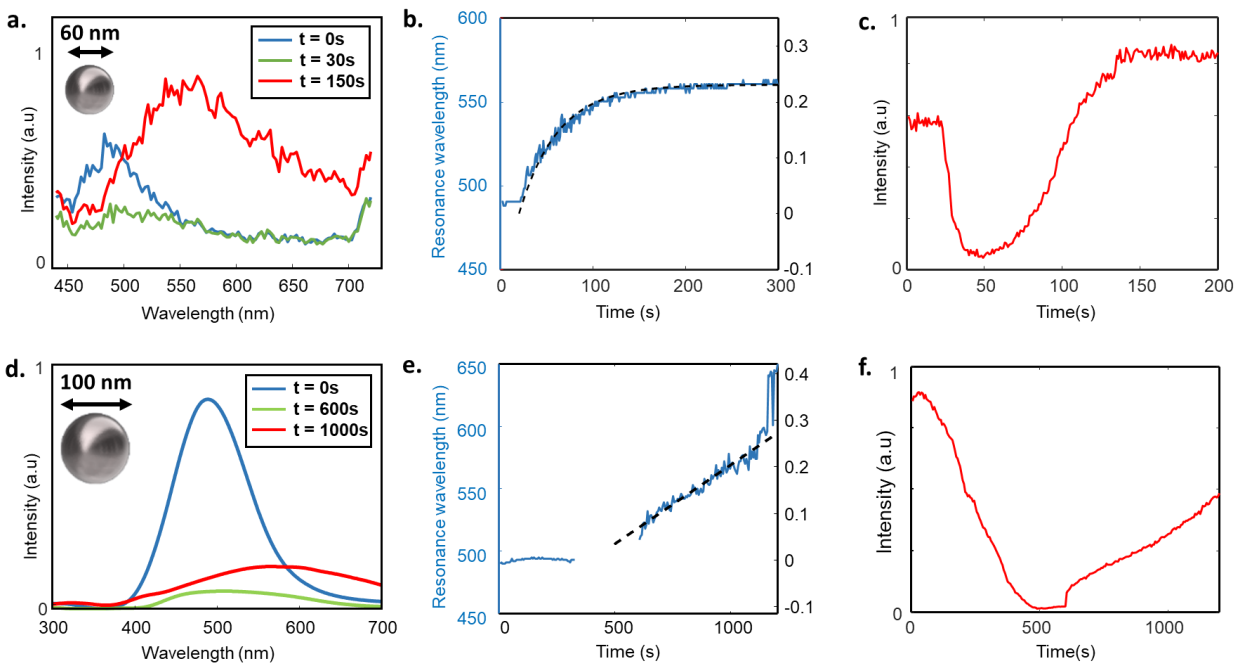

**Figure 2**: Spectral variations of 60 nm (a-c, top) and 100 nm (d-f, bottom) immobilized Ag NPs undergoing galvanic exchange owing to electrogeneration of $AuCl_4^-$ ions (triggered at t=0s). (a, c) Sample spectra at various stages of the reaction. (b, e) Variations of the spectral position of the peak and (c, f) of the total scattered intensity along the galvanic exchange process.

Two typical experiments are presented in Figure 2 for respectively a 60 and 100 nm Ag NP situated ca. 1 - 2 mm away from the Au wire electrode. In this 2D approach, a single particle is monitored within a 50x50 μm$^2$ region of interest (ROI) illuminated by the fibered white Xe lamp. The light scattered within a confocal volume (1.8x1.8x6.6 µm$^3$) centered on the particle is directed towards a spectrometer (Andor Shamrock 303i) based on an EMCCD camera (Andor iXon3 385) triggered at 1 spectrum/s. The acquisition and the anodic dissolution of the Au wire (submitted to a potential 1.1 V) are both triggered at t=0. Figure 2 shows spectra acquired at different instants, the variations of the spectral position of the peak, and the scattered intensity for 60 nm (a, b, c) and 100nm (d, e, f) Ag NPs. Note that the strong difference in the scattering cross-sections of these particles (which varies as the 6$^{th}$ power of the NP diameter in this size range) explains the stark difference in Signal-to-Noise Ratios (SNR).

For both NP sizes, changes in the scattered intensity and spectral shape start >30s after the electrochemical triggering, corresponding (as discussed below) to the time needed for the diffusive transport of AuCl$_4^-$ ions to the FoV. The transformation of the NP is first transcribed as a decrease in the scattered intensity (Figure 2c and 2f), followed by a spectral shift towards the red and an increase of the intensity (Figure 2d and 2e).

The kinetics of the reaction, which spreads over hundreds of seconds, is overall much slower than in the case of direct injection of gold ions in the solution. It is comparable to, or slower than the kinetics of the non cooperative transformation associated to AgCl precipitation observed by Jain et al.,[16,54] who reported a ca. 100-300 s transformation time of 40 nm Ag NPs while a 5 µM Au ions solution in presence of 10 mM Cl$^-$ was delivered over the NPs in the microfluidic system. This suggests that a comparable or more dilute Au ions solution is produced here by electrogeneration.

The gradual spectrum redshift is quantitatively supported by a Mie theory calculation of the scattering properties of the model nanostructures that could be formed by the galvanic exchange reaction. This optical model is presented in section SI 2. Briefly, it considers two extreme scenarii of core-shell structures, presented in Figure 3 for a 100 nm NP, in which AgCl is either present in the shell (made of a mixture of Au and AgCl, while the core is pure Ag) or the core (then made of a mixture of Ag and AgCl, while the shell is of pure Au). The optical properties of such model nanostructures are computed for different NP composition, varying the molar fraction of converted Ag metal atoms, $\delta = 1 - N_{Ag}/N_{Ag0}$, with $N_{Ag}$ and $N_{Ag0}$ the number of Ag metal atom in the converted and pristine Ag NPs respectively.

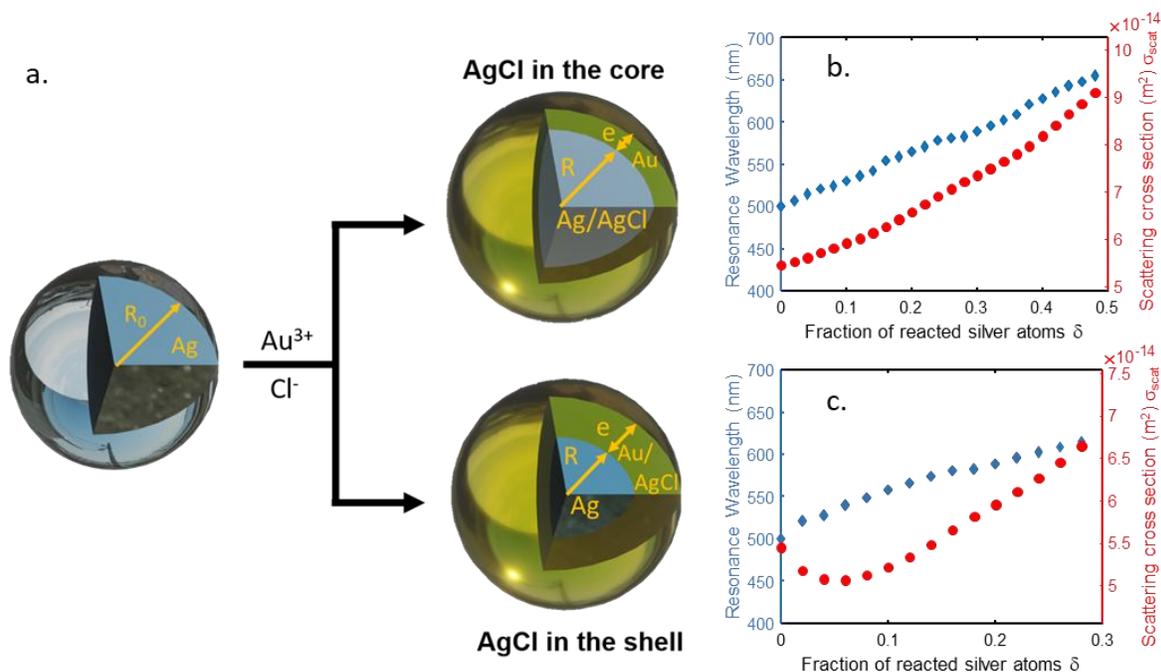

**Figure 3**: a) Ag NPs transformation pathways with two models of core-shell structures. Calculated (see SI2 for details) variations of the resonance wavelength and scattering cross section of a 100 nm Ag particle during its galvanic transformation into a Ag/AgCl core- Au shell (b) or into a Ag core- Au/AgCl shell nanoparticle (c).

The models suggest that the NP conversion into AgCl and Au can indeed be associated to the experimentally observed optical signatures variations. Interestingly, the computed nanostructure resonance wavelength $\lambda_R$ (maximum of the scattering spectrum for a 100 nm NP, blue diamonds in Figs 3b and 3c) varies linearly with the converted NP molar fraction, $\delta$,

$$\lambda_R(\delta) = \sigma\, \delta + \lambda_{R,0} \qquad (5)$$

with $\lambda_{R,0}$ the resonance wavelength of the pristine Ag NP and $\sigma$ the slope of the molar fraction dependence of the wavelength. For moderate NP conversion rates, this slope $\sigma$ is relatively independent of the conversion scenario, with values of 3.1 or 3.8 nm per % of conversion respectively for 60 and 100nm initial Ag NPs. This linear dependence thus provides a convenient way to deduce the galvanic exchange kinetics from experimental measurements of the resonance wavelength (Fig. 2b and 2e).

For the largest 100 nm Ag NP (Fig. 2e), after an induction time, the resonance wavelength $\lambda_R$ linearly increases with the reaction time, t. After ca. 1000 s of reaction, the ca. 100 nm redshift in $\lambda_R$ suggests that 26% of the NP has been converted, according to the optical model shown in fig.3. It indicates that only the outermost 5nm of the NP (10% of its radius) has been converted. Such small amount of converted material proceeds through an apparent first order kinetics described by:

$$dN_{Ag}/dt = -k_c N_{Ag} \qquad (6)$$

and yielding for the conversion $\delta$, in the limit of small $\delta$ values

$$\delta(t) = (1 - e^{-k_c t}) \approx k_c\, t \qquad (7)$$

The resonance wavelength variation with time, should then follow:

$$\lambda_R(t) = \sigma\, \delta(t) + \lambda_{R,0} \approx \sigma\, k_c\, t + \lambda_{R,0} \qquad (8)$$

from which an estimate of the conversion rate $k_c = 3\ 10^{-4}\ s^{-1}$ ensues. For the 100 nm Ag NP (composed of $3\times10^7$ Ag atoms), this corresponds to a conversion of approximately $10^4$ Ag atom per second, suggesting from the estimate of the diffusive flux of incoming $AuCl_4^-$ ions from eq.

3, i.e. of the order of $10^6$ $AuCl_4^-$ ions per second, that the galvanic exchange is rather limited by the solid phase conversion kinetics.

For the 60 nm Ag NP (Fig. 2b), the position of the spectral peak and the scattered intensity reach a plateau at t>150 s, indicating that the transformation process reaches equilibrium. The ca. 70 nm spectral redshift suggests that the reaction stops after 22% of the Ag NP has been converted, corresponding to the transformation of 2.5 nm of the outer layer of the Ag NP, a thickness in the same range as that observed for the 100 nm NP. The end of the conversion process may be caused by the precipitation of AgCl within the structure, likely intercalated between an Au shell and an Ag core. Its poor charge transport properties impede the transport and reaction of Au ions with Ag atoms. In the 100 nm Ag NP, this transformation process therefore occurs over much longer times than in the 60 nm, as clearly observed in Figure 3.

While this study validates the electrochemically-controlled delivery of dilute $AuCl_4^-$ ions, as well as the dynamic spectroscopic monitoring on static NPs, it also clearly indicates that transport mechanisms in and around the NP are crucial to the galvanic transformation mechanism. In this context, studying surface-attached NPs is a clear limitation, as the substrate can mechanically hinder ion access, and affect their interaction with the NP. It is therefore crucial to be able to study the galvanic exchange occurring in solution, on freely moving colloidal silver NPs.

**Spectroscopic study of moving particles**

To address this question, the spectroscopic system described above is associated to an adaptive optical system able to move the confocal spectroscopic collection volume in 3D and follow the random motion of a single NP. As detailed in the Methods/Experimental section, the position of the NP is determined in 3D and in real time using a holographic system. A $\lambda = 532$ nm single mode laser focused with a low numerical aperture illuminates the 50 µm x 50 µm FoV (P ≈ 30 mW at the sample, ensuring minimal thermal or photochemical effects). This illuminating beam sent through a prism is reflected off the last glass-air interface, and only light scattered by the particles is collected by the microscope objective. A dichroic mirror, directs the $\lambda = 532$ nm light towards an EMCCD camera (Andor Zyla, used at 10 frames/s) where it interferes with a "reference" collimated beam coming from the same laser to create a hologram (see Figure 1). After reconstruction of the hologram, the selected NP is superlocalized in 3D[12,59] in real time. Simultaneously, white light scattered within the confocal volume (except in the narrow band around 532 nm used for localization) is collected and directed towards the spectrograph (1 spectrum/s). The (x, y, z) position of the NP determined using holography is used to position the collection volume at the NP location, using 2 galvanometric mirrors (x, y) and a tunable lens (z). Two prerequisites should be statistically fulfilled to allow single NP tracking, i) dilution should ensure that only one NP is present in the (1.8x1.8x6.6 µm$^3$) confocal spectroscopy volume, and ii) Brownian movement must be slow enough to keep the particle within this volume during the 100 ms separating consecutive 3D localization-and-correction events.

Colloidal 100nm Ag NPs were diluted to $2.10^{-7}$ NP/mL so as to allow typically 2-3 NPs within the FoV and minimize the risk of having two NPs within the collection volume. In order to slow the Brownian motion, a 60/40 v/v mixture of a 25 mM HCl aqueous solution and glycerol was used. The viscosity of the solution is estimated as 4 mPa.s, slowing down species diffusion (ions and NPs) by a factor of 4 as compared to water.[62]

The laser beam of the holographic system was then aligned in the solution, in the middle of the microfluidic chamber (instead of the glass surface, for static NPs), and laterally off-centered 1 to 1.5 mm away from the Au electrode to avoid specular reflections on the metal, as detailed earlier. Once a NP was detected within this FoV, the electrodissolution of the Au electrode was triggered at t = 0. The first set of continuous optical monitoring was then started at t=180 s, a sufficient time, based on the static NPs conversions study, for the Au ions to diffuse towards the FoV and for NP conversion to be initiated. Owing to our current restrictions in computing buffers for the 3D optical data handling and storage, the optical monitoring of the FoV could only be performed continuously during 50s periods, each separated by data storage lapses of ca. 180 s. Based on a modeling of the temporal evolution of the $AuCl_4^-$ concentration in the FoV for the experimental electrochemical current transient (see discussion in SI 1 and Figure SI1c and SI1d), the FoV is expected to contain sub-µM $AuCl_4^-$ (increasing from 0.08 to 0.3 µM at ca. 1.25 mm from the Au electrode, see Figure SI1i) during this first monitoring interval, from t=180 s to t=230 s. During each measurement period, the 3D trajectory of the NP gives access to its hydrodynamic size $d_h$ via its Mean Square Displacement (MSD), while spectroscopy gives an insight into its composition and geometry.

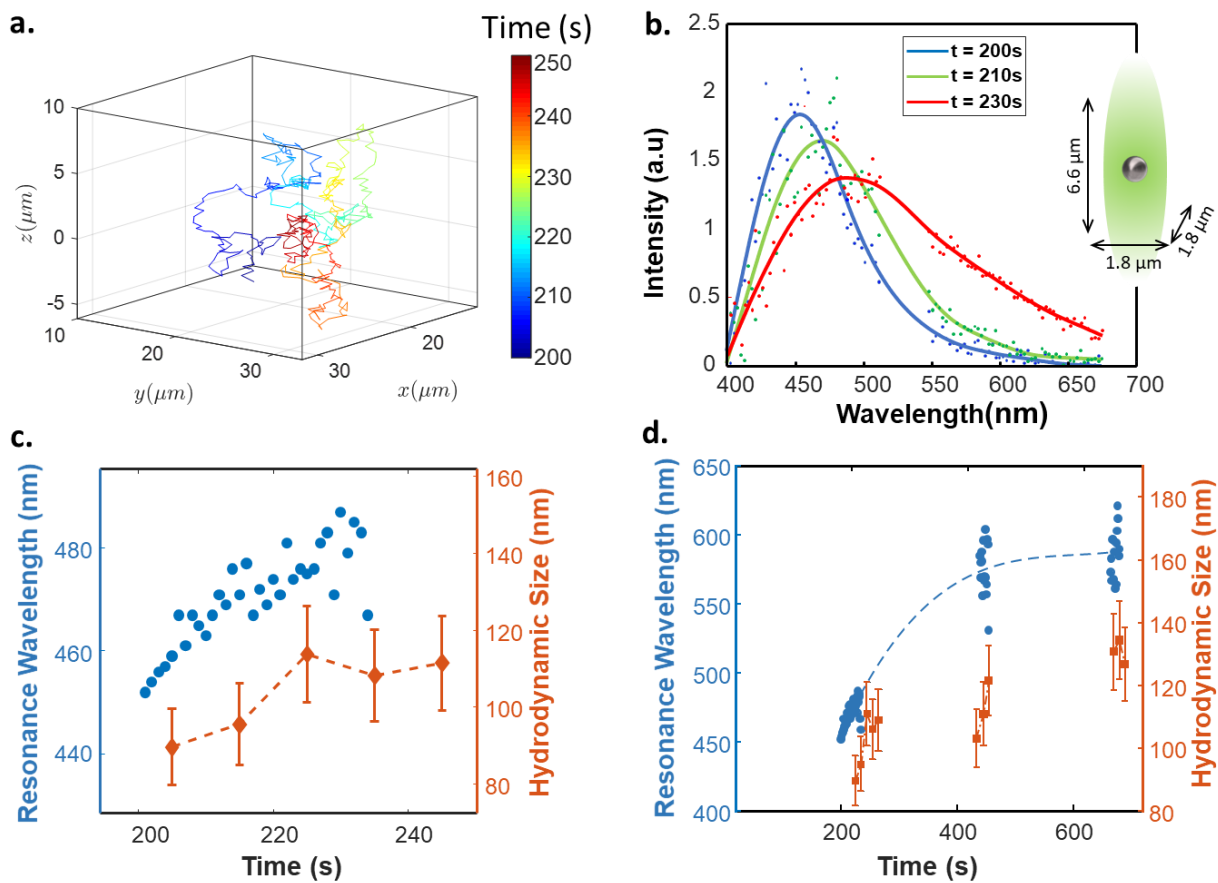

**Figure 4**: Galvanic exchange reaction on a freely moving 100nm Ag NP. a) 3D trajectory of a NP superlocalized by holographic microscopy. b) Spectra of the NP at various instants during the reaction. The inset depicts the size of the confocal collection volume (NP not to scale). c) Variation of the resonance wavelength (blue dots) and hydrodynamic radius of the NP, as determined by MSD analysis of the trajectory over 5 trajectory segments (orange dashed line). d) Monitoring of the resonance wavelength (blue) and hydrodynamic radius over 700 s (t= 200 to 250 s: same data as c; the dashed line is a guide for the eye).

One such trajectory is shown in Figure 4a. In order to derive variations of the hydrodynamic radius, $d_h$, of the NP as the galvanic exchange unfolds, this trajectory was divided into 5 independent segments of 100 points, over which a MSD analysis was conducted (MSD analysis detailed in SI 1). As shown in Figure 4c (orange dashed line), the hydrodynamic size of the NP increases from 90 to 110 nm. This is consistent with 2D studies reporting a 25 % Ag NP size increase during galvanic exchange in Cl$^-$ solution as Ag is transformed during the replacement reaction by lower-density AgCl.[54]

As observed for immobilized NPs, the scattering spectra of the NP undergoes a clear red-shift during the first set of acquisitions, shown in Figure 4 b-c, which is also a clear qualitative indication of size increase.

At longer times, as shown in Figure 4d, a clear shift of the NP spectrum towards the red is also observed between the 1$^{st}$ and 2$^{nd}$ exploration periods (separated by 230 s), suggesting the NP conversion has continued. The 3$^{rd}$ exploration period yields a $\lambda_{res} \approx 590$ nm resonance wavelength which remains mostly unchanged with respect to the previous period, revealing, as in static NPs, the limited extent of the conversion and that the transformation has reached an equilibrium.

Using the complementary spectroscopic and size variations acquired during the 1$^{st}$ period of exploration (Figure 4c), a quantitative analysis of the NP conversion kinetics can be performed. During these 30 s, both the NP size and resonance wavelength display a relatively linear behavior with time. From equations (6) and (8), the variations of resonance wavelength, $\Delta\lambda_R$, provides an estimate of the Ag NP conversion variation, $\Delta\delta$, during the exploration period, and from:

$$k_{c,3D} = \frac{\Delta\delta}{\Delta t} = \frac{1}{\sigma}\frac{\Delta\lambda_R}{\Delta t} \qquad (9)$$

an apparent conversion rate, $k_{c,3D} \approx 30/(380 \times 30) = 2.6 \, 10^{-3}$ s$^{-1}$ is estimated from the variations in Figure 4c. This corresponds to a conversion of ca. 8 $10^4$ Ag atom per second. A freely moving NP encounters various concentrations as it travels across the electrogenerated gradient. However, the 2 mm NP-electrode distance only changes by 1 % as a particle travels over 20 μm. According to the simulations of SI1, this corresponds to a maximum concentration variation of 10%. From the modelled $AuCl_4^-$ concentration profiles, an average of 0.2 μM can be adopted within the investigated FoV. This suggests a diffusive flux of $3\ 10^4$ $AuCl_4^-$ ions per second and therefore, since the Ag conversion rate is 3 times larger than that of Au deposition (eq. 4), a diffusion-limited NP conversion rate of $9\ 10^4$ Ag atom per second is expected. This value is comparable to the conversion rate evaluated by single NP spectroscopy. Although the NP conversion is slow, owing to the electrochemically controlled delivery of sub-μM Au ions, the NP conversion seems to be limited here by the diffusion of the reacting $AuCl_4^-$ ions towards the NP.

Overall, as for static NPs, the galvanic exchange in the presence of $Cl^-$ ions does not result in a full conversion of the NP. During the 1$^{st}$ exploration period, less than 9% of the NP was converted, and even for t > 400 s, when the reaction is apparently terminated from a spectroscopic point of view, a total of 30% of the NP only has been converted.

It may be delicate to relate the Ag conversion dynamics to the NP size variations. Indeed, the galvanic exchange reaction usually relies on template conservation in the AgNP. In the absence of $Cl^-$ ions, the exchange yields an Au-decorated nanocage with a size comparable to that of the pristine Ag NP. When $Cl^-$ are added, it was suggested that the template inheritance is respected but, owing to the deposition of AgCl, significantly larger nanostructures are produced. The overall NP size dynamic obtained through our *in situ* MSD monitoring is consistent with the progressive thickening of the AgCl deposit. The correlation between the conversion and size variations suggests that the conversion of Ag into AgCl follows a kinetic pattern (first order kinetics) comparable to that of the Ag consumption (6). It then suggests a proportionality between the variations in NP volume, $\Delta V_{NP}$, and conversion, $\Delta \delta$, so that $\Delta V_{NP} = \alpha \Delta \delta V_{NP,0}$, where $V_{NP,0}$ is the initial volume of the NP (or equivalently $d_{NP,0}$ its initial size) and $\alpha$ a dimensionless constant. In the limit of small $\delta$ values, a first-order expansion of the NP volume results in NP size variations correlated to the resonance wavelength variations:

$$\Delta d_{NP}/d_{NP,0} \approx \alpha \Delta \delta/3 \approx \alpha \Delta \lambda_R/3\sigma \qquad (10)$$

As shown in Figure 4d, the size expansion of the NP parallels the resonance shift over the whole galvanic exchange reaction. Equation (10) suggests that the overall 30% $\Delta\delta$ estimated from the spectra evolution would be associated to a 33% NP size increase (final NP size of $\approx$ 120 nm), in reasonable agreement with the 130 nm final hydrodynamic size obtained from the MSD analysis of our real time 3D motion monitoring.

**CONCLUSIONS**

A 3D optical and spectroscopic platform is proposed to track dynamically and in situ the reactive trajectories of individual Brownian NPs in solution. It is used here to study the galvanic exchange between gold, $AuCl_4^-$, ions and silver, Ag, metal in a chloride-containing solution at the level of individual Ag NPs either immobilized or freely moving in solution. In order to capture the whole reaction trajectory, the NP conversion dynamics is slowed down by a controlled delivery of dilute solution of the reacting $AuCl_4^-$. This is provided from (i) the *in situ* electrogeneration of $AuCl_4^-$ ions from the electrochemical oxidative dissolution of a Au wire electrode which is (ii) held 1 to 2 mm away from the optical field of view, FoV. The modelling of the experimental electrochemical current suggests that µM to sub-µM Au ions are populating the FoV, allowing the detection of the Ag NP conversion with sub-monolayer Ag atom sensitivity and a kinetic analysis of the NP conversion from the variation of its surface plasmon resonance spectrum. Particularly, the spectral variations, due to the transition of NP composition and morphology, were supported from a Mie's theory modeling of the scattering spectrum of the NP during its conversion into a core/shell nanoparticle composed of Au, AgCl and Ag. The spectroscopic analysis suggests that the conversion of the NP is halted before the full conversion of the Ag atoms, owing to the ion diffusion barrier of the AgCl deposit formed between the Au shell and Ag metals. While tracking the 3D trajectory of the moving NP undergoing the galvanic exchange, MSD analysis suggests the NP size increases, which correlates with the spectroscopic-inferred conversion of the NP and supports the formation of AgCl deposit within the nanostructure. Importantly, the transformation kinetics estimated on tethered ($k_c \approx 3 \; 10^{-4} \; s^{-1}$) and freely moving particles ($k_{c,3D} \approx 2.6 \; 10^{-3} \; s^{-1}$) differ by one order of magnitude. These values are strongly affected by uncertainties, particularly those regarding the electrode-NP distance $d$ which influences the ion concentration experienced by the NP. However, the slower transformation of surface-attached particles clearly points at a hindered ion access in the presence of the surface, as opposed to a fully accessible, freely-moving particle.

These results show that the dual monitoring of the trajectory and spectrum of single particles provides rich insight into the reaction dynamics of objects previously accessible only collectively (thus erasing the specific behavior of individual particles), or fixed on a substrate (which deeply modifies the particle-solution interaction). With further automation, monitoring larger numbers of individual particles should soon become possible, allowing to bridge the meso-scale gap between ensemble NP chemical transformation studies and the monitoring of the transformation of individual NPs proposed here. Since the holographic localization method relies on coherent optical scattering, however, it is limited to objects with scattering cross sections larger than $\sigma_{scat} \approx 10^{-16} \; cm^2$ [63]. While this gives access to a broad range of nanostructures and particles, the method is not yet sensitive enough to detect and track single molecules. In the case of light-emitting objects, however, methods able to measure the optical phase of non-coherent fluorescence emission to deduce an accurate 3D position have recently emerged [64]. Experiments extending the concept of a simultaneous tracking and spectroscopy presented here to single fluorescent molecules, or to single fluorescence-tagged entities should soon become possible.

**METHODS / EXPERIMENTAL**

**Holographic 3D localization**

As shown in Fig. 1, the experimental system is a combination of i) a holography-based particle-tracking system with ii) a confocal spectrometer. Two sources of illuminations are used in the setup: a single-longitudinal mode Nd:YAG CW laser ($\lambda = 532$nm) for holography and a white light source (Xe arc light) for the spectrometer.

The holographic tracking system[27,59] is illuminated by a single-longitudinal mode Nd:YAG CW laser ($\lambda = 532$nm), which is split in two beams, the *object* and the *reference* beam. In order to detect the weak scattering by nanoparticles and avoid blinding the camera with the excitation light, the *object beam* is sent through a glass prism into the microfluidic chamber, and reflected at the last glass/air interface. Therefore, the objective collects only the light scattered by particles, on a dark background (Olympus IX73, objective Olympus 60x NA = 0.7), and sends it towards a camera (Andor sCMOS Zyla 5.5).

The *reference beam* is sent directly to the camera as a plane wave forming an angle of ~1.5° with respect to the optical axis of the microscope. Since the object and reference waves are coherent, their interference on the camera leads to a fringe pattern, the hologram. A numerical reconstruction using the angular spectrum method[68,69] allows the reconstruction of 3D images of the light scattered in and around the plane of focusing. A region of interest comprising 16x16x8 voxels is manually chosen around the image of the targeted particle. A program determines the center of mass of the optical pattern within the region of interest [12,60], which corresponds to the 3D position of the particle. The reconstruction and 3D localization processes are time-consuming and were coded in Cuda Language, to be computed on a GPU, to obtain a 10x to 100x improvement in computing time as compared to classical CPU computing. This system can provide, in real time and up to 25 Hz (and typically 10 Hz in most of the experiments shown here) a set of coordinates (x(t), y(t), z(t)) for the position of the particle (and, optionally, a measurement of the scattering at the holographic wavelength 532 nm).

**Nanoparticle movement correction and spectroscopy**

The spectroscopic arm of the system is illuminated by a fibered Xe-arc white light source, focused with a low numerical aperture to illuminate the whole investigated region (not re-presented on Figure 1 for clarity). Again, this beam is sent to the sample through the prism and the microfluidic chamber, at an angle of 45°, to obtain total reflection at the last glass-air interface and collect only scattered white light on a dark background. Several nanoparticles are illuminated at any given time within the FoV, and averaging scattered light would lead to an ensemble response. Here, we use the entrance of an optical fiber placed in a chosen image plane as a confocal hole defining the collection volume (optical fiber of 50μm giving a collection volume of about 1.8×1.8×6.6 μm$^3$ in the object plane), chosen to enclose one nanoparticle only. However, as discussed above, if the fiber entrance and its conjugate in the sample are kept fixed, Brownian motion will quickly cause the nanoparticle to exit the collection volume. In order to compensate the lateral Brownian motion of the nanoparticle, two galvanometric mirrors (*Cambridge Technology*) are placed in a conjugate pupil of the microscope to laterally redirect a chosen 2D region of the plane of focus towards the entrance of the fiber.

In much the same way, longitudinal movements of the nanoparticle, i.e. along the optical axis of the microscope, induce defocusing of its image at the entrance of the collection fiber. These movements are corrected using a tunable lens with computer-driven focal length (*Optotune*) to maintain optimal focusing at the entrance.

Both the transverse and longitudinal movements are driven in real time by the (x, y, z) position provided by the digital holography system, and converted by a dedicated software (LabView language, with Cuda-coded holographic reconstruction and localization subroutines) into a set of voltages ($V_x$, $V_y$, $V_z$) fed to the galvanometric mirrors and tunable lens. The one-to-one relationship between (x, y, z) and ($V_x$, $V_y$, $V_z$) is determined by a calibration procedure using the positioning of a gold nanoparticle, fixed on a glass substrate, at several known positions in 3 dimensions.

The light collected by the input fiber is directed towards a spectrometer (Andor Shamrock and Ixon camera) to obtain spectra of a chosen moving nanoparticle. Owing to 3D position correction, the spectral acquisition time can be entirely decorrelated from the holographic acquisition time, and chosen to optimize either speed or spectrum signal-to-noise ratio.

**Measuring nanoparticle size with MSD**

This system thus provides dual -position and spectrum- information on the individual nanoparticle tracked holographically within the confocal detection volume of the optical fiber. The set of ($x_i$, $y_i$, $z_i$) positions allows mean-squared displacement analyses, which can provide estimates of the hydrodynamic radius of the individual nanoparticle. For 3D movement, the mean-squared displacement over a lapse $\Delta t$ is expressed as:

$$\langle r^2 \rangle = \frac{1}{N}\sum_i^N (r_i - r_{i-1})^2 = 6D\Delta t \tag{2}$$

where $D$ is the diffusion coefficient of the particle in solution and $r_i = \sqrt{x_i^2 + y_i^2 + z_i^2}$. A single set of ($x_i$, $y_i$, $z_i$) can therefore be decomposed in various $\Delta t$ lapses to plot <$r^2$> as a function of $\Delta t$ and obtain a measurement of $D$ with a linear fit. The particle diameter $d_{NP}$ can then be derived using the Stokes-Einstein equation :

$$D = \frac{k_B T}{3 \pi \eta\, d_{NP}} = \frac{\langle r^2 \rangle}{6 \Delta t} \Longrightarrow d_{NP} = \frac{2 k_B T \Delta t}{\pi \eta \langle r^2 \rangle} \tag{3}$$

**Sample preparation**

Experiments were conducted in custom-made microfluidic chamber fabricated by heat-sealing a glass slide and a cover-slip with two rings of parafilm enclosing two long, thin, pipette tips. These tips are used to inject solutions and allow air outlet, but also to insert electrodes into the chamber. In our case, the working electrode and the counter electrode are gold and silver wires, respectively.

Freely moving nanoparticles were prepared by diluting 60 and 100 nm silver nanoparticles (Sigma Aldrich) so as to observe typically $10^8$ particles/mL and minimize the risk of having two nanoparticles within the collection volume. In order to slow Brownian motion down to rates compatible with acquisition times, a 60/40 v/v mixture of ultrapure water (resistivity of 18.2 MΩ·cm$^{-1}$) and glycerol (VWR Chemicals) was used. The viscosity of the solution is estimated as 4 mPa.s.[11] Hydrochloric acid (HCl, Thermo Fisher Scientific) was also introduced into the mixture to obtain a concentration of Cl$^-$ ions of 25mM.

**Supporting Information**

**The Supporting Information is available free of charge online on the ACS Publications website. Experimental methods, electrogeneration of Au ions and modeling of consumption speed and Au ion diffusion, optical modeling of Ag-Au core shells, additionnal data on fixed and moving nanoparticles. (pdf)Author Contributions**

The manuscript was written through contributions of all authors. All authors have given approval to the final version of the manuscript.

**Funding Sources**

This work was supported by "Agence Nationale de la Recherche" (NEOCASTIP ANR-CE09-0015-01, ELISE ANR-21-CE42-0008-04, IHU FOReSIGHT ANR-18-IAHU-01) and Région Ile de France (Gene Therm - C'Nano - DIM Nano-K 2016). M.C.N thanks Sorbonne Paris Cité for funding by a "Contrat doctoral Double Culture USPC 2016" fellowship.

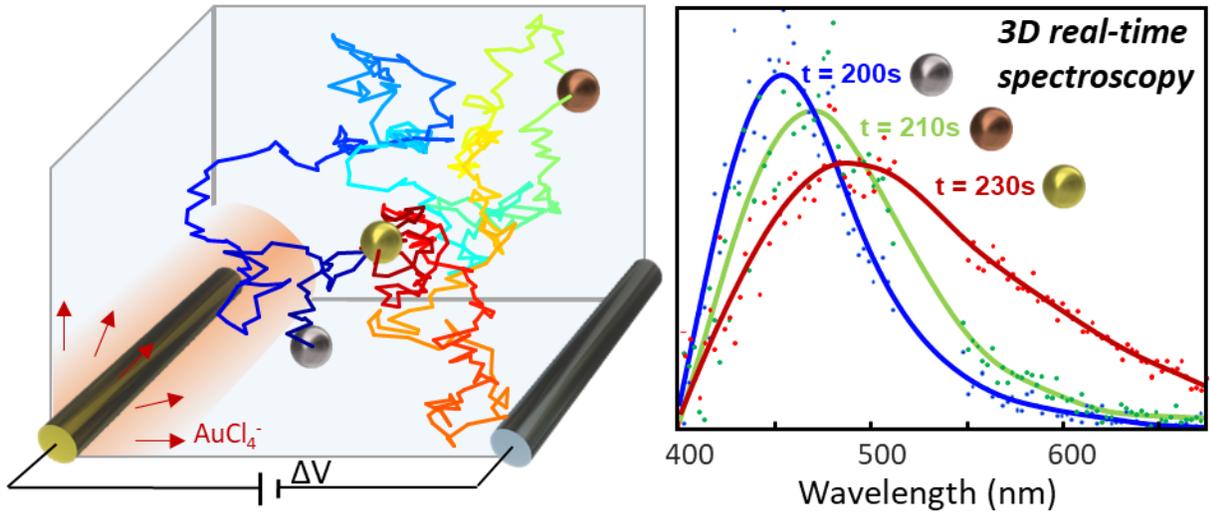